\documentclass[reprint,superscriptaddress,APL]{revtex4-1}

\pdfoutput=1
\usepackage{graphicx}
\usepackage{amsmath,bbm}

\begin{document}

\title{Nanoparticle detection in an open-access silicon microcavity}

\author{Stefan Kuhn}
\email{stefan.kuhn@univie.ac.at}
\affiliation{University of Vienna, Faculty of Physics, VCQ, Boltzmanngasse 5, 1090 Vienna, Austria}
\author{Georg Wachter}
\affiliation{University of Vienna, Faculty of Physics, VCQ, Boltzmanngasse 5, 1090 Vienna, Austria}
\affiliation{Institute for Atomic and Subatomic Physics, Vienna University of Technology, VCQ, Stadionallee 2, 1020 Vienna, Austria} 
\author{Franz-Ferdinand Wieser}
\affiliation{University of Vienna, Faculty of Physics, VCQ, Boltzmanngasse 5, 1090 Vienna, Austria}
\author{James Millen}
\affiliation{University of Vienna, Faculty of Physics, VCQ, Boltzmanngasse 5, 1090 Vienna, Austria}
\author{Michael Schneider}
\affiliation{Institute for Sensor and Actuator Systems, Vienna University of Technology, 1040 Vienna,
Austria}
\author{Johannes Schalko}
\affiliation{Institute for Sensor and Actuator Systems, Vienna University of Technology, 1040 Vienna,
Austria}
\author{Ulrich Schmid}
\affiliation{Institute for Sensor and Actuator Systems, Vienna University of Technology, 1040 Vienna,
Austria}
\author{Michael Trupke}
\affiliation{University of Vienna, Faculty of Physics, VCQ, Boltzmanngasse 5, 1090 Vienna, Austria}
\affiliation{Institute for Atomic and Subatomic Physics, Vienna University of Technology, VCQ, Stadionallee 2, 1020 Vienna, Austria} 
\author{Markus Arndt}
\affiliation{University of Vienna, Faculty of Physics, VCQ, Boltzmanngasse 5, 1090 Vienna, Austria}

\begin{abstract}
We report on the detection of free nanoparticles in a micromachined, open-access Fabry-P{\'e}rot microcavity. With a mirror separation of 130$\,\mu$m, a radius of curvature of 1.3\,mm, and a beam waist of $12\,\mu$m, the mode volume of our symmetric infrared cavity is smaller than 15\,pL. The small beam waist, together with a finesse exceeding 34,000, enables the detection of nano-scale dielectric particles in high vacuum. This device allows monitoring of the motion of individual 150\,nm radius silica nanospheres in real time. We observe strong coupling between the particles and the cavity field, a precondition for optomechanical control. We discuss the prospects for optical cooling and detection of dielectric particles smaller than $10\,$nm in radius and $1\times10^7\,$amu in mass.
\end{abstract}

\maketitle
Optical microcavities recirculate and strongly confine light, thereby enhancing the interaction between light and matter \cite{Vahala03,Zhi17}. Small mode volume cavities have been coupled to single atoms \cite{Aoki06, Trupke07}, Bose-Einstein condensates \cite{Colombe07}, organic molecules \cite{Toninelli10}, quantum dots \cite{Barbour11}, and nitrogen vacancy centers \cite{Albrecht13}. Whispering gallery mode (WGM) resonators can be used to detect and characterize label-free molecules \cite{Vollmer02}, single viruses \cite{Vollmer08}, aerosol-particles \cite{Zhu10} and particles with radii of a few tens of nanometers \cite{Lu11,Li14} when the specimen is adsorbed onto the resonator. Similar results have been achieved in photonic crystal cavities \cite{Lee07,Quan13, Liang15} and nanoplasmonic-photonic hybrid microcavities \cite{Dantham13}. 

In recent years, a growing number of research groups have utilized optical cavities to control the motion of dielectric nanoparticles \cite{Kiesel13, Asenbaum13, Millen15}.  Cavity-mediated quantum ground state cooling is predicted to be within reach for both their centre of mass motion \cite{Chang10,Romero-Isart10, Barker10} and their rotational degrees of freedom \cite{Stickler16,Hoang16}. Cold, free particles in high vacuum are considered excellent candidates for matter-wave interferometry \cite{Arndt14, Bateman14}, in a mass range where limits to established quantum theory may be explored \cite{Ghirardi90,Diosi87,Ghirardi90a,Penrose96}. The coupling of nanoparticles to a cavity field can be increased by exploiting the shape-enhanced polarizability of non-spherical particles \cite{Kuhn15, Kosloff16}, and by reducing the mode volume of the cavity. Bulk optical cavities are typically limited to beam waist radii larger than $\sim50\,\mu$m, due to limitations on the radii of curvature of the cavity mirrors. 

Here, we present the detection of free nanoparticles in transit through a chip-based, high finesse, open-access silicon Fabry-P{\'e}rot microcavity, with a 15\,pL mode volume. We detect the motion of the particles via the transmitted cavity light, and extract their velocity. Such gas-phase detection and characterization is advantageous in many fields, such as aerosol physics, nanoparticle synthesis and nanoparticulate exposure studies. In addition, we observe strong coupling between a nanoparticle and the optical field, which is of utmost importance for efficient optomechanical cooling.

\begin{figure}[]
	 {\includegraphics[width=0.46\textwidth]{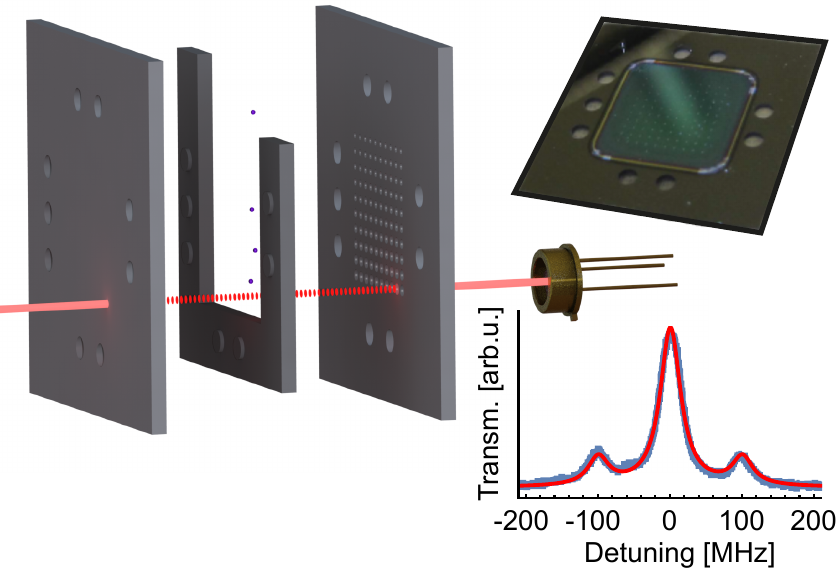}}	
\caption{\label{fig:setup} 
Schematic of the microcavity array with nanoparticle injection. The fundamental mode of a single microcavity is strongly pumped by a laser at 1547\,nm, and the transmitted light is monitored with an InGaAs photodetector. Silica nanoparticles are launched through the cavity mode via Laser Induced Acoustic Desorption \cite{Kuhn15,Millen16}. The upper inset shows a photograph of one of the mirror chips. 
The lower inset shows a scan over the TEM$_{00}$ cavity mode in transmission (blue), in which two side-bands, 100\,MHz separated from the carrier, are used to calibrate the frequency scan. From a Lorentzian fit (red) of the cavity transmission we find the cavity decay rate $\kappa/{2\pi} = (17\pm0.7)$\,MHz and deduce a Finesse of 34,000 from the ratio of free spectral range to cavity linewidth $\kappa/\pi$. }
\end{figure}

The method for fabricating and characterizing the silicon microcavities will be explained in detail elsewhere. Briefly, two silicon chips, which are the mirror image of each other, are patterned with an array of 100 mirrors with radii of curvature ranging from $70\,\mu$m to 1.4\,mm. They are separated by a $106\,\mu$m thick, silicon spacer, as illustrated in Fig.~\ref{fig:setup}, which lithographically pre-aligns the cavities. The total length of the individual cavities depends in addition on the depth of the mirrors. The mirror chips are coated with a high reflectivity, multilayer dielectric coating (Advanced Thin Films, target transmission of 15\,ppm). A single cavity, exhibiting sufficiently high finesse, is pumped with 25\,mW at a wavelength of 1547\,nm (Toptica CTL-1550) to excite a TEM$_{00}$ mode, and the laser frequency is stabilized to the cavity resonance via side-of-fringe locking using the back-reflected beam. Here, we use an optical cavity with a length $L = (130\pm3)\,\mu$m, a free spectral range $\nu_{\rm{FSR}} = c/2L = (1.15\pm0.03)\,$THz, a mode coupling of 30$\,\%$, mirror radius of curvature $R = (1.3\pm0.2)\,$mm, a mode waist (radius) $w_0 = (12\pm1)\,\mu$m, a decay rate $\kappa/{2\pi} = (17\pm0.7)$\,MHz, and a finesse of $\mathcal{F} = 34,000\pm1,300$. This yields a mode volume $V_m = 14.5\,$pL or $3,900\,\lambda^3$, and an on-resonance intracavity intensity at the waist of approximately $2\times10^{7}\,$Wcm$^{-2}$. The cavity finesse is limited by the deviation of the mirror shape from an ideal parabolic profile, and can be improved by optimization of the fabrication parameters \cite{Kleckner2010}. 

Silica nanoparticles (Bangs Laboratories) of radius $r = (150\pm20)\,$nm are launched through the cavity field at a pressure of $10^{-7}$\,mbar, using Laser Induced Acoustic Desorption \cite{Kuhn15, Millen16}. To detect the nanoparticles, the cavity input light is detuned from resonance by $\Delta = -2.3\,\kappa$, and the transmitted light is monitored on a photodiode. The presence of a dielectric particle inside the cavity mode effectively increases the optical path length, thus shifting the cavity towards resonance and increasing the amount of light transmitted through the mirrors. The particle also scatters light out of the cavity mode, decreasing the amount of light transmitted. A net increase of the transmitted signal is a clear sign of strong, dispersive coupling between the particle and the cavity field \cite{Aspelmeyer14}. This allows the detection of particles with a signal-to-noise ratio (SNR) of more than 35, enabling detection of silica particles as small as 50\,nm in radius.

\begin{figure}[]
	 {\includegraphics[width=0.46\textwidth]{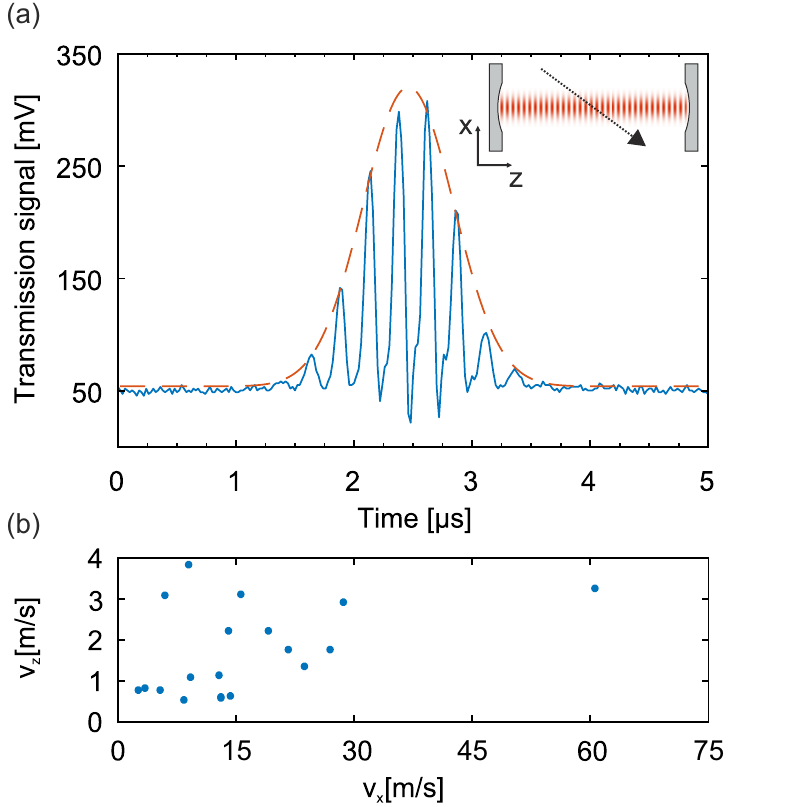}}	
\caption{\label{fig:results} 
(a) An example of the microcavity transmission as a 150\,nm radius silica nanoparticle traverses the cavity field. The laser which pumps the cavity is red-detuned from the empty cavity resonance by $2.3\,\kappa$. The envelope due to the motion in the $x$-direction (see inset) across the cavity waist is indicated by the red dotted line. The periodic structure is due to the particle's motion in the $z$-direction, crossing multiple nodes of the cavity field. From this signal, the velocity in the $x$- and $z$-directions can be extracted ($v_x = (15.6\pm 0.1)$\,ms$^{-1}$ and $v_z = (3.13\pm 0.07)$\,ms$^{-1}$). The signal drops below the baseline due to light scattering out of the cavity mode. (b) The extracted transverse velocity $v_z$ and forward velocity $v_x$ for 19 particles.}
\end{figure}

An example nanoparticle transit is shown in Fig.~\ref{fig:results}a. As the particle traverses the optical mode in the $x$-direction i.e.\ perpendicular to the optical axis of the cavity, the transmitted signal increases with a Gaussian envelope (red dotted line), directly mirroring the Gaussian waist $w_0$ of the microcavity mode. Since $w_0$ is known from the cavity geometry, we can extract the velocity in the $x$-direction, $v_x = (15.6\pm 0.1)$\,ms$^{-1}$. There is also a fast modulation of the signal, as the nanoparticle crosses the optical standing-wave in the $z$-direction. Since the wavelength of the light is known, we can extract the velocity in the $z$-direction, $v_z = (3.13\pm 0.07)$\,ms$^{-1}$. There is a slight decrease in signal below the baseline level in Fig.~\ref{fig:results}a, which is due to the particle scattering light from the cavity mode. In total, the particle dispersively shifts the cavity resonance by more than $2\,\kappa$, confirming strong coupling between the particle and the cavity field. The extracted velocities for 19 nanoparticles are shown in Fig.~\ref{fig:results}b.

\begin{figure*}[]
	 {\includegraphics[width=0.95\textwidth]{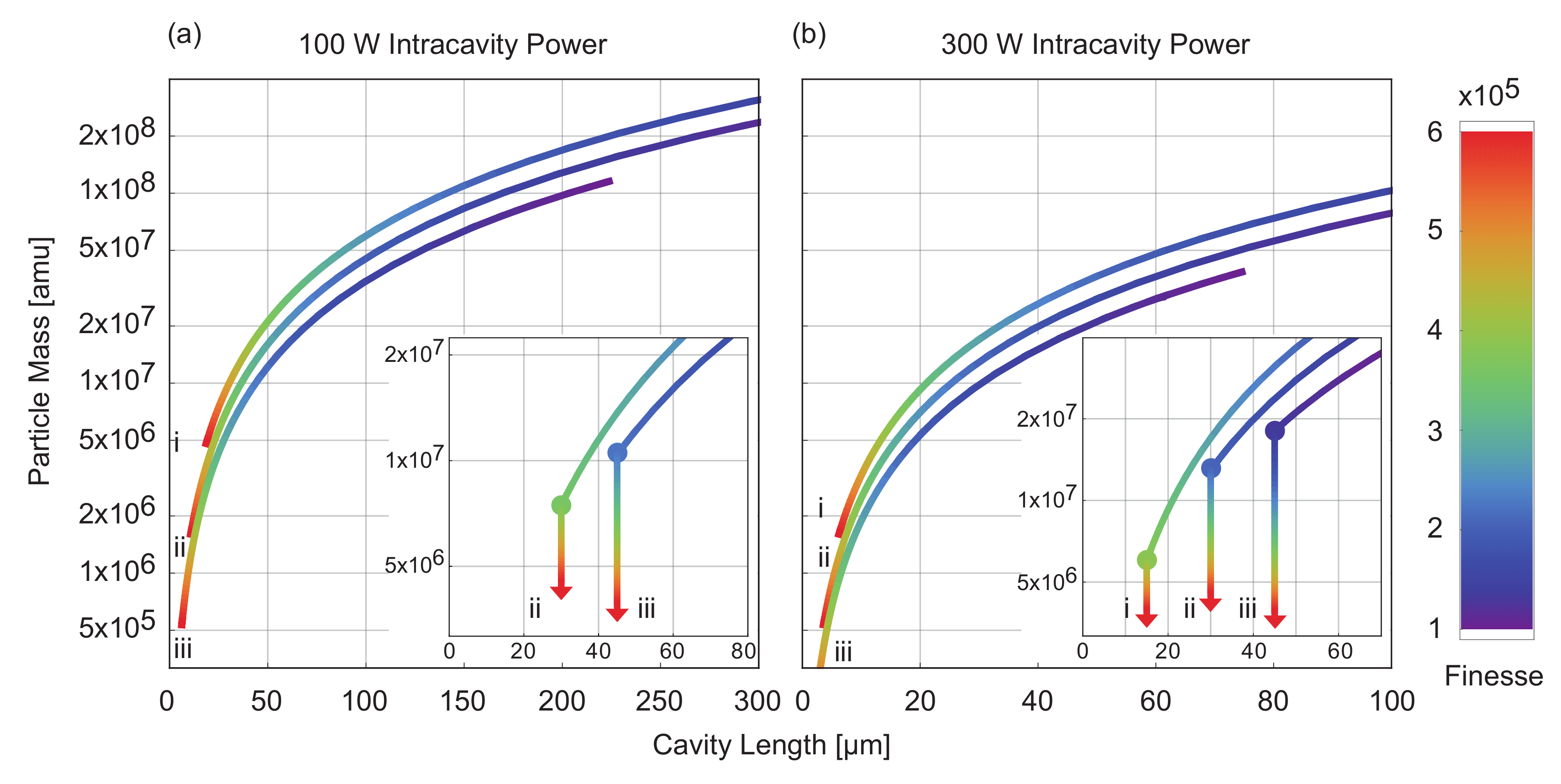}}	
\caption{\label{fig:cooling} 
Optimizing microcavity parameters to achieve optomechanical control of low-mass silicon nanospheres. We find the required cavity parameters to enable cooling of particles of a given mass under strong coupling and resolved sideband conditions (see text). We consider an intra-cavity power of (a) 100\,W and (b) 300\,W, at fixed $L/R$ ratios of i) 0.5, ii) 1.0 and iii) 1.5. The insets illustrate that, even with a lower bound of $R = 30\,\mu$m (large points), it is still possible to reduce the coolable mass by increasing the cavity finesse. For realistic cavity parameters and geometries, cooling of silicon spheres with masses below $1\times10^7\,$amu, corresponding to a silicon sphere of radius below 12\,nm, is feasible.}
\end{figure*}

We now consider one specific application of such microcavities, namely their use for the optomechanical cooling of dielectric nanoparticles \cite{Horak97,Vuletic00,Chang10,Romero-Isart10,Barker10,Kiesel13,Asenbaum13,Millen15}. Cooling is believed to be necessary, for instance, to enable matter-wave interferometry with nanoscale objects \cite{Arndt14, Bateman14}. In particular, we will consider cooling of silicon nanoparticles due to their favourable dielectric properties \cite{Asenbaum13}, their applicability for nanofabrication \cite{Kosloff16} and their compatibility with optical ionization gratings \cite{Reiger06}. Such experiments are challenging at high mass, since the short de Broglie wavelength of a massive object requires a long interferometer. For a given resolution in the interference pattern on the detector screen, the required flight time through the interferometer scales linearly with the particle mass \cite{Note1}, limiting the mass to the $10^6 - 10^7$\,amu range \cite{Bateman14} i.e.\ silicon spheres of 6-12\,nm radius. Such small particles cannot be cooled in macroscopic cavities.

For optimal cooling the following criteria must be met: A) Operating in the regime of strong coupling, i.e. the dispersive frequency shift of the cavity resonance induced by the particle 
\begin{align}
U_0 = \frac{2 \pi \omega_L r^3}{V_m} \frac{\varepsilon - 1}{\varepsilon + 2},
\end{align} 
(laser frequency $\omega_L$ and relative permittivity of the particle $\varepsilon$) is larger than the cavity decay rate $\kappa = c \pi/2\mathcal{F} L$. B) Working in the resolved side-band limit to ensure that the response of the cavity field amplitude is retarded relative to the motion of the particle. Therefore, the axial mechanical trapping frequency of the particle 

\begin{align}
\omega_z = \sqrt{\frac{24 k^2 P_{\rm{cav}}}{\pi w_0^2 \rho c} \frac{\varepsilon - 1}{\varepsilon + 2}},
\end{align} 
(wave-vector of the cavity field $k$, intra-cavity power $P_{\rm{cav}}$ and particle density $\rho$) needs to be larger than $\kappa$. In order to cool small masses, the following cavity parameters can be optimized: finesse $\mathcal{F}$, beam waist radius 

\begin{align}
w_0=\sqrt{\frac{\lambda}{2 \pi}\sqrt{L(2R-L)}},
\end{align} 
via $L$ and $R$, and the intra-cavity power $P_{\rm{cav}}$. 

The parameter-space for optimal cooling is displayed in Fig.~\ref{fig:cooling}. The coloring of the curves indicates the minimum finesse, for a given ratio $L/R$ and intra-cavity power, required to fulfill conditions A \& B. This finesse, in turn, sets the minimal mass that can be cooled for the given set of geometrical constraints. Increasing the mass does not allow for a reduction in finesse due to condition B. At a given $L/R$, however, the minimum mass can be decreased by improving the finesse, as shown in the insets of Fig.~\ref{fig:cooling}. We consider fixed ratios $L/R$ of i) 0.5, ii) 1.0 \cite{Note2}, and iii) 1.5, and intra-cavity powers of a) 100\,W and b) 300\,W. Hence, by moving to smaller cavities with higher finesse \cite{Derntl14} ($\mathcal{F} > 2\times10^5$, $R\approx20\,\mu$m, $L\approx20-30\,\mu$m, $P_{\rm{cav}}>300\,$W) it will be possible to cool silicon nanoparticles with masses below $1\times10^7\,$amu, corresponding to a sphere of radius 12\,nm. Assuming the same SNR as presented here, optimized microcavities could detect particles down to 5\,nm in radius.   

In conclusion, we present the detection of free nanoparticles in an open-access silicon microcavity. We observe their transit via the transmitted cavity light, extract their velocity, and observe strong coupling between the particle and intracavity field. Such a system will be useful for optomechanics, and for characterization and detection of nanoparticles, bio-molecules, viruses, and aerosols. With further improvements, this microcavity system will enable cavity cooling of a diverse range of sub\,-10\,nm particles, which are suitable for matter-wave interferometry in a hitherto unexplored mass range of $10^6 - 10^7\,$amu. 

\vspace{3mm}
\noindent \textbf{Acknowledgments} 	\\
\noindent We are grateful for financial support by the Austrian Science Fund (FWF) through the projects P27297, ``SiC-EiC'', DK-CoQuS (W1210) and DK-Solids4Fun (W1243). We further acknowledge funding from the Vienna University of Technology research funds. JM acknowledges funding from the European Union's Horizon 2020 research and innovation programme under the Marie Sk{\l}odowska-Curie grant agreement No 654532.

\end{document}